\documentclass{ws-ijmpa}
\usepackage{amsmath}
\usepackage{graphicx}
\def\a{\alpha}

\def\b{\beta}

\def\p{\partial}
\def\s{\sigma}

\def\g{\gamma}

\def\De{\Delta}
\def\ov{\overline}
\def\ld{\lambda}
\def\Ld{\Lambda}

\def\e{\eta}

\def\om{\omega}

\def\rh{\rho}

\def\b{\beta}

\def\a{\alpha}
\def\ast{\alpha^*}

\def\pdellx'{\frac{\partial}{\partial x'}}
\def\pdellw'{\frac{\partial}{\partial w'}}

\newcommand{\be}{\begin{equation}}
\newcommand{\ee}{\end{equation}}
\def\bed{\begin{displaymath}}
\def\eed{\end{displaymath}}
\def\bea{\begin{eqnarray}}
\def\eea{\end{eqncrray}}
\def\[{$$}
\def\]{$$}
\begin{document}
\title{Big Jets Model with CPT Invariance and Dynamics of Expansion with Quantum Yang-Mills Gravity}
\author{
Jong-Ping Hsu\\
 Department of Physics, 
 University of Massachusetts Dartmouth,\\ 
North Dartmouth, MA 02747, USA\\
Leonardo Hsu\\
Department of Chemistry and Physics,  Santa Rosa Junior College,\\
Santa Rosa, CA 95401, USA\\
Daniel Katz\\
 Department of Physics, University of Massachusetts Lowell,\\
Lowell, MA 01854, USA\\} 

\date{April 2018}
\maketitle
{\small   Based on particle physics, the fundamental CPT invariance suggests a Big Jets model for the beginning of the universe, in which two oppositely directed jets evolved into a gigantic `matter half-universe' and a gigantic `antimatter half-universe' after annihilation and decay processes. In the geometric-optics limit, quantum Yang-Mills gravity with $T_4$ translational gauge symmetry in flat spacetime leads to an effective metric tensor in the Hamilton-Jacobi equation for macroscopic objects.  This effective metric tensor does not exist in the wave equations of quantum particles.   For cosmological expansion, we assume that an ``effective metric tensor" for spacetime geometry based on Yang-Mills gravity corresponds to the usual FLRW form.  Dynamical equations of expansion for the matter half-universe are obtained and solved.  The time-dependent scale factors and the estimated age of the universes, $t^{YM}_o \approx 15.3 \times 10^9 yr$, based on Yang-Mills gravity are consistent with experiments.  CPT invariance implies that the same evolution process and dynamics of cosmic expansion also hold for the distant `antimatter half-universe.'}

\bigskip
Keywords: CPT invariance; big jets model; Yang-Mills gravity; particle cosmology; dynamics of expansion.

PACS numbers:  11.15.-q,  98.80.Bp, 98.80.Es 

\bigskip

\bigskip

In order to describe the complicated motion and distribution of matter and energy across
the cosmos, we employ an effective metric tensor for the spacetime geometry of the observable portion of the expanding universe.  The FLRW model with the metric tensor $g_{\mu\nu}=(1,-a^2,-a^2,-a^2)$ was developed to play the role of the physical metric of expanding space, where the Robertson-Walker scale factor $a=a(t)$  is a function of time and denotes the change of distance between astronomical objects.\cite{1}  For example, the Friedman equations lead to the solutions $a(t) \propto t^{2/3}$ for a matter dominated universe.

Recently, a theory of quantum Yang-Mills gravity was developed on the basis of the translational $T_4$ gauge symmetry in flat spacetime.\cite{2,3,4}  In contrast to the usual gauge theories with an arbitrary (Lorentz) scalar gauge function, the translational $T_4$ symmetry in inertial frames involves arbitrary vector gauge functions $\Ld^{\mu}(x)$ with a constraint $\p_\mu\Ld^\mu(x)=0$.  Yang-Mills gravity is consistent with experiments and brings gravity back to the arena of gauge field theory and quantum mechanics.  It also provides a solution to difficulties in physics such as the incompatibility between `Einstein's principle of general coordinate invariance and all the modern schemes for  quantum mechanical description of nature.'\cite{5}. Moreover, all our discussions are based on inertial frames, in which the space and time coordinates have their usual well-defined operational meaning.  

Quantum Yang-Mills gravity reveals that the apparent curvature of spacetime appears to be a manifestation of the flat spacetime translational gauge symmetry for the wave equations of quantum particles in the geometric-optics limit.\cite{3} Thus, according to quantum Yang-Mills gravity, macroscopic objects moves as if they were in a curved spacetime because they are described by the Hamilton-Jacobi type equation, $G^{\mu\nu}(\p_\mu S)(\p_\nu S)-m^2=0$,\footnote{This classical equation derived from the quantum Yang-Mills gravity in flat spacetime in the geometric-optics limit was called `Einstein-Grossmann equation' of motion in memory of their collaboration.  Such an equation is crucial for Yang-Mills gravity to be consistent with experiments.} with an `effective Riemann metric tensor' $G^{\mu\nu}$, which is a function of $T_4$ gauge fields.

With our limited knowledge of the universe at present, it is premature to conclude that we have observed the most part of the universe.\cite{6}   In particular, the Big Bang picture of the universe has difficulties with the established laws of particle physics, such as the CPT invariance\cite{7} or matter-antimatter symmetry.  It is natural to postulate Big Jets model with CPT invariance for the birth of the universe based on particle cosmology.  In analogy to the high energy interactions of particles, in which many jets could be produced, we postulate that the universe is created similar to the simplest two Big Jets with incredible amount of energies, involving baryons, leptons and gauge bosons.  It is also natural to assume that the fundamental (dynamic and geometric) symmetry principles in particle physics, such as color $SU_3$ and 4-dimensional spacetime symmetry, were there to dictate the ensuing processes of annihilation and decay of particles and the evolution of these particles through various interactions.  One may roughly picture the Big Jets model as two gigantic fireballs moving away in opposite directions, each with an unimaginable number of particles and antiparticles.  After annihilations and decays, one fireball happened to be dominated by stable particles (baryons, electrons, antineutrinos) and formed matter half-universe, while the other was dominated by stable antiparticles (antibaryons, positrons, neutrinos) and formed antimatter half-universe.\cite{8}  From the vantage viewpoint of observers in each gigantic fireball, the evolution process is similar to that of a hot Big Bang model. 

In particle physics, the combination of local quantum fields, Lorentz invariance and the spin-statistics relations for quantizations of fields implies the exact CPT invariance,\cite{7} where C (charge conjugate) denotes changing the sign of a charge, P (Parity, ${\bf r}\to -{\bf r}$) denotes space inversion and T denotes time reversal ($t \to -t$).  This exact CPT invariance assures exact lifetime and mass equalities between particles and antiparticles.  It also implies opposite electroweak (and chromo-) interaction properties between particles and antiparticles, and the same gravitational interactions between particles and antiparticles (in the sense of local quantum Yang-Mills gravity based on flat spacetime).  Thus, CPT invariance leads to a maximum symmetry between matter and antimatter in the whole universe and suggests Big Jets Model. 

Since the matter fireball has expanded and cooled down to become a `matter blackbody' at about 2.7K, the antimatter fireball should have also expanded and cooled down to become an 'antimatter blackbody' at about 2.7K, according to CPT invariance.  Presumably, these two blackbodies are separated by an extremely large distance (probably much much more than 14 billion light years) such that the lights emitted from, say, individual anti-supernovae and anti-galaxies are too weak to be detected by us.  Nevertheless, the microwaves emitted from the antimatter blackbody as a whole may be detected.  If the matter blackbody could be modeled as a spherical blackbody, then the Big Jets model predicts that its hemisphere that faces the the antimatter blackbody should be warmer than the opposite hemisphere, which will have a uniform temperature of 2.7K.  This prediction of a `semi-dipole anisotropy' could be tested by satellite CMB data.\cite{8}  The usual Big Bang model would not result in such a `semi-dipole asymmetry' of temperature in the CMB data.

 Based on the requirements of homogeneity and isotropy on large scales,  the FLRW model  assumes
 \be
 ds^2=g_{\mu\nu}dx^\mu dx^\nu, \ \ \ \  g_{\mu\nu}=\left(1,-a^2(t),-a^2(t),-a^2(t)\right), 
 \ee
 $$
 T_{\mu\nu}=\left(\rh(t), a^2(t)p(t), a^2(t)p(t), a^2(t)p(t)\right), \ \ \ \  c=\hbar=1.
 $$
 where $T_{\mu\nu}$ is the energy-momentum tensor for macroscopic bodies.\cite{1}  The metric tensor $g_{\mu\nu}(t)$ holds in a general framework.  The expansion of the universe, as described by the scale factor $a(t)$ in (1), is governed by the gravitational field equation in general relativity (GR).
 
 Since quantum Yang-Mills gravity is based on particle physics in flat spacetime, it is natural to employ Yang-Mills gravity with translational gauge symmetry in inertial frames to discuss the dynamics of expanding matter half-universe, to determine the expansion rate and to estimate the age of the universe.   The Big Jets model with CPT invariance in particle physics suggests that the effective expanding spacetime of the `antimatter half-universe' should be treated similarly and separately.  In this connection, it is reasonable to assume that the interaction between `matter half-universe' and `antimatter half-universe' has negligible effects on their individual expansions, since they are separated by an extremely large distance.   
  
 Within the expanding `matter half-universe' (mhu) with homogeneity and isotropy, it is natural to explore the implications of quantum Yang-Mills gravity on the effective covariant metric tensor $G^{mhu}_{\mu\nu}$, which is directly related to the line element $ds^2$ of the expanding space-time geometry,
 \be
ds^2=G^{mhu}_{\mu\nu}dx^\mu dx^\nu, \ \ \ \ \  G_{\mu\nu}^{mhu}=\left(B^2(t),-A^2(t), -A^2(t),-A^2(t)\right).
\ee
Similar to FLRW model,\cite{1} we assume the effective tensor $G^{mhu}_{\mu\nu}$ in (2) to take the diagonal form.

According to quantum Yang-Mills gravity based on a flat spacetime, classical macroscopic objects moves as if they were in a curved spacetime because they obey a relativistic Hamilton-Jacobi type equation with an effective Riemann metric tensor in the geometric-optics limit.\cite{3} In the present (HHK) model of cosmic expansion for the matter half-universe, quantum Yang-Mills gravity in inertial frames and the cosmological principle  imply the following simple relations\cite{2}
\be
  G^{mhu}_{\mu\nu} = \e^{\a\b} J_{\a\mu}(t) J_{\b\nu}(t), \ \ \ \  \ \  \e^{\a\b}=(1,-1,-1,-1),
\ee
\be
    J_{\mu\nu}(t)=(B(t),A(t),A(t),A(t)), 
\ee
In Yang-Mills gravity, the `effective covariant metric tensors'  $G^{mhu}_{\mu\nu}$ in (3) and (4) are governed by the $T_4$ gauge invariant gravitational field equations in inertial frames,
\be
 g^2 S_{\mu\nu}=\p_\ld\left(J_{\ld'\rh'} C_{\rh\mu\nu}\e^{\rh\rh'}-J_{\ld'\a'} C_{\a\b}^{~~~\b}\e^{\a\a'} \e_{\mu\nu} + C_{\mu\b}^{~~~\b}J_{\nu\ld'}\right)\e^{\ld\ld'}
\ee
$$
-C_{\mu\a}^{~~~\b}\p_\nu J_{\a'\b}\e^{\a\a'} + C_{\mu\b}^{~~~\b}\p_\nu J_{\a\a'}\e^{\a\a'} - C_{\ld\b}^{~~~\b}\p_\nu J_{\mu\ld'}\e^{\ld\ld'},  
$$
\be
C_{\mu\nu\a}=J_{\mu\ld}(\p_{\ld'} J_{\nu\a})\e^{\ld\ld'} - J_{\nu\ld}(\p_{\ld'} J_{\mu\a})\e^{\ld\ld'}, \ \ \   C_{\mu\a}^{~~~\b}=C_{\mu\a\b'}\e^{\b\b'}
\ee
where $g^2=8\pi G$ and $\mu$ and $\nu$ should be made symmetric in the field equation (5).\cite{2}   $C_{\mu\nu\a}$ is the $T_4$ gauge curvature and $S_{\mu\nu}=\ov{\psi}i\g_\mu \De_\nu \psi - i(\De_\nu \ov{\psi})\g_\mu \psi$ is the energy-momentum of coupled fermions with $\De_\nu=J_{\nu\ld} \p^\ld$ and $\p^\mu S_{\mu\nu}=0$.

  Based on (4) and (6), the non-vanishing components of the $T_4$ gauge curvature $C_{\mu\nu\a}$ are
\be
C_{0ik}= - C_{i0k}= - B {\dot A} \e_{ik}, \ \ \ \ \ \ \    \dot{A}=dA(t)/dt.
\ee
All other components vanish.

In analogy to (1) in the FLRW model, we assume that the `matter half-universe' is homogeneous and isotropic on the Hubble scale, as observed in inertial frames.  Thus, the energy-momentum tensor of the `matter half-universe' everywhere takes a simple diagonal form:
\be
T_{\mu\nu}= \left(\rh(t) B^2(t),-P(t) A^2(t),-P(t) A^2(t),-P(t) A^2(t)\right),
\ee
where $\rh(t)$ and $P(t)$ are the energy density and pressure of macroscopic bodies, respectively.  This is related to the universal gravitational interaction of the $T_4$ tensor fields. 
The effective tensors $G^{mhu}_{\mu\nu}$ and $J_{\mu\nu}$ in (2)-(4), together with the replacement of $S_{\mu\nu}$  in  (5) by $ T_{\mu\nu}$ in (8), are postulated in the HHK model.\footnote{In the HHK model, one has $T_\mu^\nu=(\rh,P,P,P)$ and $T_{\mu\nu}=T_\mu^\ld G^{mhu}_{\ld\nu}$.}  Thus, the dynamics of expansion is  dictated by the Yang-Mills gravity, which exercises the full power of its  $T_4$ gauge symmetry to govern the scale factors $A(t)$ and $B(t)$ in (4) and, hence, the effective metric tensor $G^{mhu}_{\mu\nu}$ in (2) and (3) for the expansion of the matter half-universe.

The Yang-Mills gravitational field equation (5), together with (3)-(4) and (6)-(8), leads to two independent equations
\be
 6 B {\dot A}^2=g^2 \rh B^2, 
\ee
$$
 -2(B^2 {\ddot A} + 2B {\dot B} {\dot A}) = -g^2 P A^2,    
$$
corresponding to $g^2 T_{00}$ and $g^2 T_  {11}$ respectively. The other two components $g^2 T_  {22}$ and $g^2 T_  {33}$ give the same equation as $g^2 T_  {11}$, i.e., the second equation in (9).  This result is due to the special time dependence of $ G^{mhu}_{\mu\nu}$ and $ J_{\mu\nu}$ in (3)-(4), and the simplicity of the $T_4$ gauge curvature $C_{\mu\nu\a}$ as shown in (7).

Using the equation of state $P=\om \rh$ with constant $\om$ and the usual expanding relation of the energy density $\rh =\rh_o/ A^{3}(t)$,  (9) leads to the following equations for the spacetime expansion with the scale factors $A(t)$ and $B(t)$,
\be
\left( 5A \ddot A + 6{\dot A}^2\right) A^6 {\dot A}^4=\frac{(g^2 \rh_o)^3\om}{72}, \ \ \ \ \ \ \  B=\frac{6{\dot A}^2 A^3}{g^2 \rh_o}.
\ee
These are the basic equations governing the expansion of the matter half-universe, according to Yang-Mills gravity.
For a `matter dominated' (md) cosmos, we use $A(t) =\a t^n$ with constant $\a$ and $n$ in (10) to obtain the solutions for $A(t)$ and $B(t)$,
 \be
A(t)_{md} = \a t^{1/2}, \ \ \ \ \   \a=\left(\frac{8 g^6 \om \rh_o^3}{9}  \right)^{1/12},   
\ee
$$
B(t)_{md}=\b t^{1/2}, \ \ \ \ \ \ \  \b=\frac{3\a^5}{2 g^2 \rh_o}.
$$
For a radiation dominated (rd) cosmos, in which $\om=1/3$ and $\rh = \rh_o/ A^{4}$,\cite{1} (the extra factor of $1/A(t)$ in the energy density $\rh$ is related to the red-shift  of the radiation wavelength as the cosmos expands), the time-dependent scale factors $A(t)$ and $B(t)$ are found to be
\be
A(t)_{rd} = \a' t^{2/5}, \ \ \ \ \   \a'=\left(\frac{5^6 g^6 \om \rh_o^3}{2^6 \times 36}  \right)^{1/15},  
 \ee
$$
B(t)_{rd}=\b' t^{2/5}, \ \ \ \ \ \ \  \b'=\frac{12 \a'^6}{25 g^2 \rh_o}.
$$
Thus, the effective covariant metric tensor $G^{mhu}_{\mu\nu}$ in (2) and (3) and the spacetime geometry of the expanding matter half-universe are determined by the spacetime translational gauge symmetry of the gravitational interaction. 

For comparison, the scale factor $a(t)$ in the FLRW model is given by $a(t)\propto t^{2/3}$ for matter dominated cosmos and $a(t) \propto t^{1/2}$ for radiation dominated cosmos.\cite{1}  Thus, the results of Yang-Mills gravity in (11) and (12) lead to slower expansions than those based on GR and give a different age of the universe.

Let us estimate the age of the universe $t^{YM}_o$ based on Yang-Mills gravity in flat spacetime with the scale factors in (11) for a matter dominated cosmos.  Since there is only one expansion rate $H(t)=\dot{A}(t)/A(t)$, it is natural to have the relation $\dot{X}(t) \propto H(t)$, which holds in any inertial frame with the Cartesian coordinates $(X,Y,Z)$.  Because of the isotropy, it suffices to consider only the $X$ coordinate, without loss of generality.  Furthermore, there is only one natural  length scale involved in the problem, i.e., the present Hubble length $X_o=cH_o^{-1}$.  Thus,  the dimensional analysis suggests the following equation for the expanding matter half-universe,
\be
\dot{X}(t)=H(t) X_o, \ \ \ \ \  H(t)=\dot{A}(t)/A(t).
\ee
This relation enables us to compare the age of the universe for different models with different expansion rates $H(t)$.  For the a matter dominated cosmos, let us write $A(t) = \a t^{\s}$.  The solutions to (13) are
\be
\frac{1}{X_o} \int _{X_a}^{X_b} dX =  ln\left(\frac{t_b}{t_a}\right)^\s,
\ee
$$
 \ \ \ \ \   \frac{t_b}{t_a}=\frac{t_o}{t_a} = e^{1/\s}, \ \ \ \ \ \ \  X_b - X_a=X_o=cH_o^{-1}, \ \ \  e=2.718,   
  $$
where $X_a$ and $t_a$ are some suitable small numbers such that when the universe expands to the present size, i.e., $X_b - X_a=X_o= c/H_o$, the time $t_b$ is naturally identified as the age of the universe, $t_o=t_b$.  

 For our calculations, the values of $t_a$ and $X_a$ are not important because the results for the relation of two ages of the universe in two models with different scale functions are independent of them.  It is the ratios of the values of $A(t)$ at different times that are important.  For a matter dominated universe, let us compare the age of the universe $t_o^{{GR}}$ based on GR (in the Einstein-de Sitter model\cite{1} with the scale factor $a(t)=a_o t^{2/3}$) and  the age $t_o^{{YM}}$ based on Yang-Mills gravity (in the present HHK model with the scale factor $A(t)=\a t^{1/2}$):
 \be
 \frac{t^{YM}_o}{t_a} = e^{2}, \ \ \ \ \  H^{YM}(t)=\frac{\dot{A}}{A}=\frac{1}{2t}, \ \ \  {HHK\ model},
 \ee
 \be
 \frac{t^{GR}_o}{t_a} = e^{3/2}, \ \ \ \ H^{GR}(t)=\frac{\dot{a}}{a}=\frac{2}{3t}, \ \ \  {EdS \ model}.
 \ee
In the Einstein-de Sitter model, one has the age of the universe $t_o^{GR} \approx 9.3\times 10^9 yr$, which appears to be too small and difficult to reconcile
 with the measurement of stellar evolution.\cite{9,1}  It follows from (15) and (16) that the `age of the universe' $t^{YM}_o$ according to Yang-Mills gravity is approximately
\be
t^{YM}_o = \sqrt{e} \ t_o^{GR} \approx 15.3 \times 10^9 yr,
\ee
 which is consistent with experiments.  The approximate result (17) for the age of the universe is based on Yang-Mills gravity and is interpreted to be started from the birth of the two Big Jets.\cite{10,11}  For a better result, the calculation should take into account the varying nature of the matter-energy content of the matter half-universe. 
 Nevertheless, it is reasonable to approximate that the matter half-universe has been dominated by non-relativistic matter for most of its existence.  Exact CPT invariance in HHK model implies that the result (17) holds also for the antimatter half-universe.  The value (17) appears to be reasonable because the cosmos expands slower in the HHK model than that in the Einstein-de Sitter or FLRW model,\cite{12,13} so that it needs more time or larger $t^{YM}_o$ to reach its currents Hubble length $X_o=c H_o^{-1}$.
  
We may remark that if one uses a slightly different relation $\dot{X}(t)=H(t) X(t),$\cite{9} instead of that in (13) to compare $t^{YM}_o$ and $ t_o^{GR}$.  The result for the ratio $t^{YM}_o/t_o^{GR}$ then depends on the choice of the initial time $t_a$, i.e. the lower limit of the time integration and, hence, is not reliable.   Also, the singularity at $t_a=0$ in (14) should not be taken seriously because the metric tensor and scale factors in (2) and (13) are only effective for large space and time intervals related to the motion of macroscopic bodies, according to quantum Yang-Mills gravity.  As long as the estimation does not depend on the choice of $t_a$, the result may be more reliable.    

If the energy-momentum tensor\cite{14} $T_{\mu\nu}$ were dominated by the vacuum energy-momentum,\cite{1} then $T_{\mu\nu}\approx T^V_{\mu\nu} = \rh_{_V} G^{mhu}_{\mu\nu}$, where $\rh_{_V}$ is a constant independent of spacetime position.  Thus, if one uses $T^V_{\mu\nu}$ and equations (2)-(7), one obtains $A(t) \propto B(t)\propto t^2$.\footnote{Strictly speaking, such a constant vacuum energy-momentum tensor in inertial frames (with $\e_{\mu\nu}$) is not allowed within the framework of local field theory such as quantum Yang-Mills gravity.} 

Within the HHK model of particle cosmology, conservation of momentum implies that the Big Jets explosion is located in the middle between the matter half-universe and antimatter half universe, and, hence, cannot be observed within our matter half-universe. 
Also, the spacetime framework of quantum Yang-Mills gravity and the Big Jets model is flat, so that there is no flatness problem\cite{1} in the HHK model of cosmology.  Moreover, the HHK model has interesting implications on the `effective speed of light' $C$ given by $ds^2=B_{md}^2 dt^2 - A_{md}^2 dr^2=0$ in the expanding spacetime geometry.  From (11), one obtains $C=dr/dt=\b/\a=(3\om/2)^{1/3}$, which depends on the equation of state $P=\rh \om$.  For the radiation dominated cosmos (12) with $\om=1/3$, we have the effective speed of light $C=2^{1/3}\approx 1.26$, which is close to the usual value $c=1$ in inertial frames.
 Further physical implications of the Big Jets model with CPT invariance and the effective metric tensor (2) for the spacetime geometry shall be discussed in future work.  

We stress that, within the framework of quantum Yang-Mills gravity in inertial frames, both the resultant $G_{\mu\nu}$ in the Hamilton-Jacobi type equation for the motion of macroscopic objects and the purely time-dependent $G_{\mu\nu}^{mhu}$ for expanding spacetime geometry are only effective metric tensors.   It did not escape our attention that a `cosmic standard time' for the time in (2)-(4) and the expansion rate $H(t)=\dot A(t)/A(t)$ could be physically realized by using a grid of synchronized clocks in an arbitrarily chosen inertial frame and by requiring this clock system to be used by all observers in all frames.\cite{15,16,17}  In this way, one would still have a spacetime framework with four-dimensional symmetry and a `cosmic time.'\footnote{Note that any inertial frame can be arbitrarily chosen to set up such a standard time.  Therefore, it is not unique and, hence, not absolute.  See also  pp.17-20, ref. 10.}

The exact CPT invariance in particle physics predicts the existence of a distant `antimatter half-universe,'\footnote{We may remark that the picture of matter half-universe and antimatter half-universe resembles the Ying-Yang diagram, i.e., ancient Chinese cosmological image for the concept of a primal entity.} whose cosmic expanding geometry is also described by equations (2)-(17).  Such a distant antimatter half-universe (or blackbody) could be tested by a careful analysis of the satellite CMB data to see if there is a `semi-dipole anisotropy' of $\approx 10^{-5}$ or smaller.\cite{8}
 
  The work was supported in part by the Jing Shin Research Fund of the UMassD Foundation.

\bibliographystyle{unsrt}
{
\end{document}

 One may also use a slightly different relation to compare $t^{YM}_o$ and 
 $\ t_o^{GR}$.  Suppose one uses the relation $\dot{X}(t)=H(t) X(t),$\cite{9} instead of that in (13).  Suppose one writes $H(t)=\s/t$, one obtains the 
 result $ {X_b}/{X_a}=(t_b/t_a)^\s$ with the interpretation that when $X_b=X_o$, one has $t_b=t_o$, where a subscript o denoting a present value.  Thus, one obtains a larger value, $t^{YM}_o=(t^{GR}_o)^{4/3} \approx 19.5\times 10^9 yr$ for matter dominated half-universe.  These results for the age $t^{YM}_o$ of the universe are indeed independent of the value of $t_a$ and $X_a$. (((This estimate will depends on $t_a$.???????)))